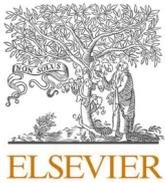

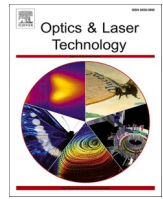

# Sensitivity-optimized strongly coupled multicore fiber-based thermometer

Josu Amorebieta [a, *], Angel Ortega-Gomez [b, *], Rubén Fernández [b], Enrique Antonio-Lopez [c], Axel Schülzgen [c], Joseba Zubia [b], Rodrigo Amezcua-Correa [c], Gaizka Durana [b], Joel Villatoro [b,d]

[a] Department of Applied Mathematics, Engineering School of Bilbao, University of the Basque Country (UPV/EHU), Plaza Ingeniero Torres Quevedo, 1, E-48013 Bilbao, Spain
[b] Department of Communications Engineering, University of the Basque Country, Bilbao 48013, Spain
[c] CREOL The College of Optics and Photonics, University of Central Florida, Orlando 162700, FL, US
[d] Ikerbasque-Basque Foundation for Science, Bilbao E-48011, Spain

## A B S T R A C T

In this paper, we report on a multicore fiber-based (MCF) temperature sensor that operates in a wide thermal range and that is robustly packaged to withstand harsh environments. To develop the sensor, the fundamentals concerning the effect of temperature on such fibers have been analyzed in detail to predict the most temperature sensitive MCF geometry. Thanks to it, the device, which operates in reflection mode and consists of a short segment of strongly coupled MCF fusion spliced to a standard single mode fiber, shows higher sensitivity than other devices with identical configuration. Regarding its packaging, it consists of an inner ceramic and two outer metallic tubes to provide rigidity and protection against impacts or dirt. The device was calibrated for a thermal range from −25 °C to 900 °C and a K-type thermocouple was used as reference. Our results suggest that the manufactured optical thermometer is as accurate as the electronic one, reaching a sensitivity up to 29.426 pm/°C with the advantage of being passive, compact and easy to fabricate and interrogate. Therefore, we believe this device is appealing for industrial applications that require highly sensitive temperature sensing in very demanding environments, and that the analysis included in this work could be analogously applied to develop sensitivity-optimized devices for other parameters of interest.

## 1. Introduction

Temperature measurement requirements have always been very demanding in diverse industrial sectors in terms of sensing conditions and extreme temperature operating ranges. Commonly, such applications require high sensitivity, as they are related to critical processes in which precise and real-time monitoring capabilities are essential for their correct operation to prevent accidents, failures or malfunctions. Moreover, usually, the environments in which such measurements are carried out are not favorable, which causes the devices to be subjected to high levels of stress that may reduce their lifetime. Representative examples of such sectors are the aeronautical industry, where the high temperatures inside turbines due to combustion need to be monitored; the petrochemical industry, where there are flammable and explosive substances that need to be maintained in certain temperature ranges; the food industry, where some products need to be preserved at sub-zero temperatures, etc. As a result, the demand for highly sensitive and robust thermometers capable to operate in wide thermal ranges has remained unaltered through years.

Nowadays, electronic thermometers are the most spread devices for industrial temperature sensing. Among them, thermocouples should be highlighted [1–6], as they are a very mature, reliable and cheap technology. However, they may not be the best candidates to deploy in environments with flammable gases or high levels of radiation. In the last decades, several optical fiber-based sensors have appeared as an alternative to overcome such limitation thanks to their passiveness and their intrinsic sensitivity to temperature. Despite the fact that first optical fiber-based thermometers had very limited thermal operating ranges, the continuous improvement in materials and structures has allowed expanding them to wider ranges [7–11], and therefore, becoming a realistic alternative to thermocouples and other temperature sensors used in industrial applications.

The most common and spread optical fiber temperature sensors are based on fiber Bragg gratings (FBGs) and Fabry-Perot interferometry (FPI). On the one hand, FBGs, and especially regenerated FBGs, offer sensitivities around 10 pm/°C, although their setups require expensive lasers and interrogators [12–17]. On the other hand, FPIs offer small size and sensitivities around 15 pm/°C, although their reproducibility may be affected by the difficulty to obtain uniform and identical cavities [18–23]. Long period gratings (LPGs) have been used as temperature sensing elements as well, but their intrinsic sensitivity to the surrounding medium forces proper isolation or compensation mechanisms to be sensitive only to temperature [24–26].

In recent years, the development of specialty optical fibers has provided new alternatives for optical fiber temperature sensing. This progress has been significant especially by the use of multicore fibers (MCFs). Briefly, MCFs consist of multiple cores embedded in a common cladding. Depending on their geometrical arrangement, two groups can be acknowledged: weakly and strongly coupled MCFs [27]. On the one hand, if the cores are far enough from each other to have weak interaction, it can be assumed that they are decoupled, as there is almost null crosstalk among them. Such fibers are mainly used for telecommunication purposes, as they allow having multiple independent channels on the same physical medium [28,29]. Weakly coupled MCFs have been used for sensing purposes as well and have shown significant performance in this matter. However, their main drawback to be used as sensors relies on the complex interrogation setups that they require in order to analyze individually the light from each of the cores [30]. On the other hand, when the cores are close enough to interact with each other, they are strongly coupled, which means there is a heavy crosstalk among them that results in a periodic coupling of light among cores. Such coupling is very sensitive to several mechanical effects on the MCF, and for this reason, strongly MCFs have been exploited for sensing applications [31].

Using MCFs as the sensitive element to develop optical thermometers provides several advantages such as ease of fabrication and interrogation, capability to withstand temperatures up to 1000 °C and sensitivities around 50 pm/°C [32–34]. However, such fibers are sensitive simultaneously to different parameters as bending, strain, curvature, etc [31,35–37]. Thus, in order to make them only sensitive to temperature and to be protected against external factors as impacts, they require proper packaging, which may have negative impact on the thermal sensitivity, as the packaging may lower the temperature transfer from the surrounding environment to the MCF [38,39]. To compensate such loss of sensitivity, a detailed analysis of the MCF fundamentals regarding the effect of temperature needs to be done to predict and select the most sensitive geometry to manufacture the optical thermometer.

Herein, we report on a ruggedized temperature sensor based on a strongly coupled MCF capable to operate in a wide thermal range from −25 °C to 900 °C. Prior to manufacturing it, MCF fundamentals regarding temperature sensitivity were studied and applied for two particular MCF geometries. According to the results provided by such analysis, the most sensitive geometry was selected to be used in the optical thermometer in order to compensate the aforementioned decrease in sensitivity due to the packaging. The device consists of a short segment of such MCF fusion spliced to the distal end of a single mode fiber (SMF). The end of MCF is cleaved to act as a low reflectivity mirror and to operate in reflection mode. The simplicity of this structure made it appropriate to evaluate the sensitivity of different MCFs straightforwardly, as the latter is highlighted in such structure as the MCF is the only sensitive element in the device. Regarding its packaging, it consists of three layers (one ceramic and two metallic tubes) and has been designed to avoid any bending or curvature on the fiber and to provide rigidity and protection against external factors as impacts or dirt. In this way, the device is extremely robust, which makes it suitable for very demanding environments at the same time that it is highly sensitive, reaching values up to 29.43 pm/°C. The reflection spectrum of the manufactured device has a unique and well-defined peak that shifts with temperature. As temperature is codified in wavelength, by tracking such maximum, the relationship between temperature and shift is easy to establish. For its calibration, a K-type thermocouple was used as a reference, as it is commonly taken as standard in the industry; and results indicate that the manufactured device is as accurate as the thermocouple. The benefits of this work rely on the conclusions derived from the analysis of MCF fundamentals regarding temperature sensitivity, which allow ruggedizing the packaging and still offer higher sensitivity than previously reported MCF optical thermometers that use the same SMF-MCF structure as the one in this work.

## 2. Operating principle, sensor design and fabrication

In strongly coupled MCFs, the cores are close enough to each other to induce the modes through each individual core to overlap among them [33]. This overlapping produces a cyclic power coupling among cores that can be described by the coupled mode theory (CMT) [40,41]. According to it, the resulting modes, which are called supermodes (SPs), are the linear combination of the modes propagating through each individual guide [42] and orthogonal among them. In the simplest case, when there are two coupled waveguides, it is possible to calculate the power transferred from one guide into the other by applying the CMT. As it can be deduced from such theory, the normalized power at a certain distance along the propagation axis $z$ in the input guide $i$ can be expressed as:

$$P_i(z) = cos^2(Sz) + cos^2(\gamma)sin^2(Sz) \tag{1}$$

In Eq. (1)' $\tan(\gamma) = k/\delta$, $S = \sqrt{\delta^2 + k^2}$, $k$ is the coupling coefficient between modes, and $\delta = (\beta_1 - \beta_2)/2$, where $\beta_1$ and $\beta_2$ are the propagation constants of the modes of each individual guide. In such equation, the key parameter that explains the transferred power between guides is $\delta$. If $\delta = 0$, optical guides are identical ($\beta_1 = \beta_2$), which implies that the power is transferred completely from one guide into the other. On the contrary, if $\delta \neq 0$, guides are not identical($\beta_1 \neq \beta_2$), which means that the power will not transfer entirely from one guide to the adjacent, causing a residual amount of power to remain always in the input guide. The understanding and application of this phenomenon is crucial to develop switches and filters, as by modifying the material or the size of the guides, the propagation constants of the modes can be adjusted in order to fabricate ad-hoc devices [43].

Particularizing Eq. (1) for SMF-MCF structures, when the fundamental $LP_{01}$ mode of the SMF is injected in the central core of the MCF, its two fundamental supermodes ($SP_{01}$ and $SP_{02}$) are excited [32,40,41,44]. Under this condition, the normalized coupled power along the propagation axis $z$ for a strongly coupled MCF as the one depicted in Fig. 1a, can be expressed as:

$$P(z) = cos^2\left(\sqrt{N+1} \ \frac{\pi\Delta n}{\lambda}z\right) + \frac{1}{N+1}sin^2\left(\sqrt{N+1} \ \frac{\pi\Delta n}{\lambda}z\right) \tag{2}$$

where $\Delta n$ is the difference between the effective refractive indices of the propagating coupled SPs, $N$ is the number of cores surrounding the central one (see Fig. 1a), and $\lambda$ is the emitted wavelength. Moreover, in Eq. (2), it is assumed that the diameter and physical properties of each core are identical and that each core supports only its fundamental mode. Regarding the number of cores, it is important to point out that Eq. (2) is valid for $N > 1$, as for the case of $N = 1$, there are two identical adjacent guides, where the power is transferred completely between both guides, and therefore the normalized coupled power can be expressed as in [34]. In Eq. (2), it can be noticed that a residual power remains in the central core, indicated by the term $[1/(N+1)]sin^2\left(\sqrt{N+1} \ \frac{\pi\Delta n}{\lambda}z\right)$, which means that in such MCFs $\delta \neq 0$. Thus, the normalized power in the central core for such MCF geometry will always oscillate in the $[1/(N+1), 1]$ range. As all the cores are identical, and therefore, have identical $\beta$, it can be assumed that the core distribution is responsible for this phenomenon [40].

Predicting the amount of power in the central core at any propagation distance and for any wavelength allows using MCFs for different applications, such as filters, switches, optical couplers with specific split ratios, etc. To design such devices, the key factor is the remaining power in the guide of interest (which is the central core in this work), as this parameter indicates the minimum achievable power in it. The methodology to obtain a specific amount of such power is based on varying $N$ (see Eq. (2)), or on modifying the ratio between the size of the central and adjacent cores [41] (ratio $a/b$, see Fig. 1a). For an application in which only certain specific fractions of power are required, modifying the number of cores is an acceptable solution. On the contrary, if any other amount of remaining power is required, changing the ratio of the size of the cores $(a/b)$ is a better solution as such methodology provides higher flexibility. However, the accuracy in the fabrication process of MCFs with such specific size ratios may be challenging.

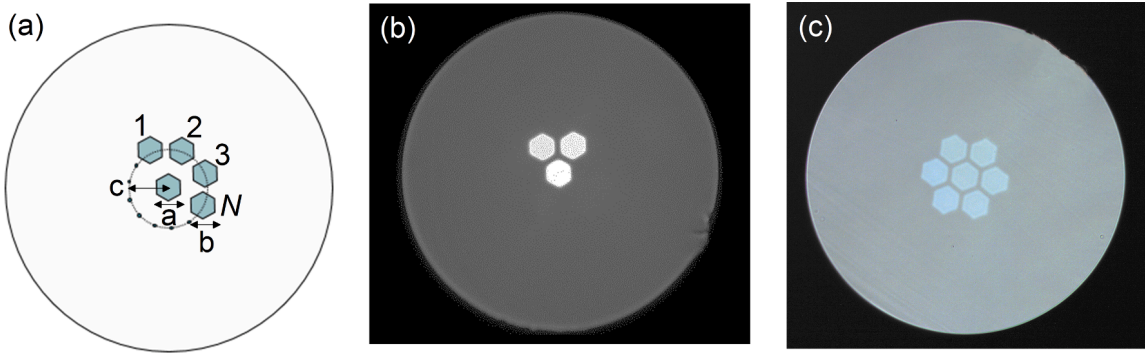

**Fig. 1.** (a) Schematic of the generic structure of the core distribution of the MCFs analyzed in this work and photographs of the particular geometries of the MCFs in this work: (b) asymmetric with respect to the horizontal axis (aMCF), (c) and with radial symmetry (sMCF).

Recently, several researches based on sensors comprised of several MCF segments fusion spliced in series have been reported [45,46]. The equation that expresses the response of such structures is the product among the response of each segment in the cascade. In this manner, the generic equation of the normalized output power of such devices is as follows:

$$P(z)_{cascade} = \prod_{l=1}^{p} P_l(z) \tag{3}$$

where $p$ is the number of MCF segments in the cascade.

In this work, the thermal sensitivity of two particular strongly coupled MCF architectures was analyzed. Such geometries have in common all of their physical properties: the germanium doped silica cores inlayed in a pure silica cladding of 125 μm of diameter, the diameter of individual cores of 9 μm, the distance among cores of 11.5 μm ($a = b = 9$ μm and $c = 11.5$ μm according to Fig. 1a), and their numerical aperture (NA) of 0.14 at 1550 nm to match that of a SMF. In addition, both MCF architectures have a core at their geometrical center, which corresponds with the one where the light is launched. The difference relies on the number of adjacent cores ($N$) and their geometrical distribution around the central one. On the one hand, the MCF in Fig. 1b is asymmetric with respect to the horizontal axis (aMCF) in such a manner that the adjacent cores are surrounding it forming an equilateral triangle. On the other hand, the MCF in Fig. 1c has seven cores, where the six adjacent cores are arranged concentrically with radial symmetry (sMCF). As the core distribution of each MCF is different, Eq. (2) has to be particularized for each case. Hence, for each of the described MCF geometries, their corresponding normalized coupled power can be expressed as:

$$P_{aMCF}(z) = cos^2\left(\sqrt{3}\ \frac{\pi\Delta n}{\lambda}z\right) + \frac{1}{3}sin^2\left(\sqrt{3}\ \frac{\pi\Delta n}{\lambda}z\right) \tag{4}$$

$$P_{sMCF}(z) = cos^2\left(\sqrt{7}\ \frac{\pi\Delta n}{\lambda}z\right) + \frac{1}{7}sin^2\left(\sqrt{7}\ \frac{\pi\Delta n}{\lambda}z\right) \tag{5}$$

When two identical (twin) segments of the aforementioned MCFs are cascaded creating a SMF-MCF-SMF-MCF-SMF structure and operating in transmission mode, the operating principle expressed in Eq. (3) has to be particularized for each MCF-based cascade:

$$P_{TwinaMCF} = cos^4\left(\sqrt{3}\ \frac{\pi\Delta n}{\lambda}L\right) + \frac{1}{9}sin^4\left(\sqrt{3}\ \frac{\pi\Delta n}{\lambda}L\right) + \frac{1}{6}sin^2\left(2\sqrt{3}\ \frac{\pi\Delta n}{\lambda}L\right) \tag{6}$$

$$P_{TwinsMCF} = cos^4\left(\sqrt{7}\ \frac{\pi\Delta n}{\lambda}L\right) + \frac{1}{49}sin^4\left(\sqrt{7}\ \frac{\pi\Delta n}{\lambda}L\right) + \frac{1}{14}sin^2\left(2\sqrt{7}\ \frac{\pi\Delta n}{\lambda}L\right) \tag{7}$$

Eq. (6) and (7) show that the cascaded configuration provides a narrower spectrum compared with a single structure (SMF-MCF), due to the cos and sin raised to the 4th in twin structures compared to the squared cos and sin in single configuration (Eq. (4) and (5)), whereas their corresponding phase remains unaltered. In addition, the amount of remaining power in the central core in any cascaded configuration is lower than that in the single one, which implies an increase in the visibility in the spectrum proportional to the number of cascaded MCF segments in the structure.

Regarding the effect of temperature on such MCFs, the measurement is absolute as it is codified in wavelength. In the resulting spectrum of the device, there is a straightforward relation between the temperature and the wavelength of the peak, ($\lambda_m$), in such a way that the later shifts with temperature proportionally. In order to determinate the position of such maximum, $\lambda_m$ has to be isolated from the generic expression of the phase in Eq. (2) or in Eq. (3), as they are identical, and equalized to $2\pi m$ (Eq. (8)), where $m$ is an integer number that determinates the tracked $\lambda_m$. Thus, there is a unique $m$ for each maximum in the spectrum, where higher numbers of such integer correspond with longer wavelengths. Furthermore, $L$ determinates the distribution of the peaks in the spectrum, which implies that a variation of $L$ causes a proportional variation of $m$ in a manner they compensate each other.

$$\lambda_m = \frac{\sqrt{N+1}}{m}\Delta nL \tag{8}$$

where $L = L_f *(1 + \alpha\Delta T)$, $L_f$ is the length of the MCF segment at room temperature ($T_r = 25\ °C$), $\alpha$ is the thermal expansion coefficient and $\Delta T$ is the variation of temperature with respect to room temperature ($T$-$T_r$).

To study and predict the wavelength shift caused by temperature changes, the partial derivative with respect to temperature has to be applied to Eq. (8), as shown in Eq. (9). As the phase remains unaltered in any MCF-based structure (comprised of single, twin, or an undefined number of cascaded elements), Eq. (8) is identical for all cases, and therefore, it can be concluded that the thermal sensitivity does not depend on the number of MCF segments in the device.

$$\frac{\partial\lambda_m}{\partial T} = \frac{L_f}{m}\sqrt{N+1}\left[\frac{\partial\Delta n}{\partial T}*(1 + \alpha\Delta T) + \Delta n\alpha\right] \tag{9}$$

From Eq. (9), it is assumed that the most significant parameters that have impact on thermal sensitivity are $\Delta n$ and $L_f$, as $\alpha$ is related to the physical properties of the material of the MCF. As it has been explained previously, such physical properties are identical for the aMCF and sMCF under study, and therefore, the value of $\alpha$ is identical for them. For that reason, the analysis of thermal sensitivity should be focused on $\Delta n$ and $L_f$. On the one hand, $\Delta n$, and therefore, $\partial\Delta n/\partial T$, are related to the SPs, and, therefore, to the geometry and physical properties of the MCF, such as the number of cores, their distribution, etc. On the other hand, $L_f$ is the length at room temperature of the MCF segment. Although it can be noticed from Eq. (9) that the thermal sensitivity is proportional to $L_f$, it can be assumed that its impact on sensitivity is negligible, as the ratio $L_f/m$ is fixed independently of the initial length of the MCF. Therefore, it can be concluded that the length of MCF segments in the sensor has negligible impact on the thermal sensitivity.

To demonstrate the aforementioned theoretical prediction, samples consisting of different lengths and configurations of bare sMCF were manufactured and exposed to several temperature cycles from 200 °C to 500 °C and back to 200 °C in steps of 100 °C, after being subjected to an annealing process to avoid hysteresis [38,47]. The fabricated and tested configurations were: 1) a sample with a sMCF segment of 12.5 mm (Fig. 2a), 2) a sample of twin cascaded sMCF segments of 12.5 mm (Fig. 2b) and 3) a sample with a sMCF segment of 25 mm (Fig. 2c). Regarding the twin structure in Fig. 2b, it consists of a SMF-MCF-SMF structure. Due to its reflection mode configuration, the cleaved end of the SMF segment acted as a mirror. By means of this configuration, the light propagates forward and backward along the structure, passing twice though the MCF segment. Thus, with only one segment of the MCF, a twin structure can be easily manufactured. The spectrum at room temperature (25 °C) of each structure is depicted in Fig. 2d.

Briefly, to fabricate such specimens, a precision fiber cleaver (Fujikura CT105) and a precision fusion splicer (Fujikura 100P +) were used. With the first, segments of the desired length with an error of ± 10 μm were obtained; whereas with the second, minimal insertion losses (below 0.1 dB) in the SMF-MCF splice may be achieved. To keep the MCF segments straight, each specimen was introduced in a ceramic tube (Omega TRX-005132–6). The inner diameter of each bore of the ceramic tubes is 127 μm, which is slightly bigger than the outer diameter of the MCF (125 μm). In this manner, the MCF fitted tightly in it, avoiding any bending or curvature, which could cause a shift in the spectrum that would be difficult to distinguish from that caused by temperature.

The interrogation setup consisted of a broadband superluminescent diode (SLED) centered at 1550 nm (Safibra, s.r.o.), a fiber optic circulator (FOC), and a commercial spectrometer (I-MON-512 High Speed, Ibsen Photonics) to analyze the reflected light. Thus, the shift of the $\lambda_m$ in the spectrum caused by temperature could be tracked with picometer accuracy. These values were correlated with temperature, which was measured with a K-type thermocouple. The devices were tested in a high temperature calibrated furnace (Fluke 9150), and the evolution of their respective spectra is shown in Fig. 3.

The results of tests are summarized in Table 1. Such results indicate that, as predicted theoretically, the initial length of the MCF ($L_f$) and cascading such elements do have negligible impact on sensitivity for temperature measurements. However, this does not imply that $L_f$ can be a disposable parameter for the design of optical MCF-based thermometer, as it is still a relevant design factor that defines the shape of the spectrum, and therefore, $\lambda_m$, as it can be acknowledged in Fig. 2d. For this type of sensors, a unique and well-defined peak is pursued, as it facilitates tracking the shift of such maximum easily and avoids any ambiguity in the measurements due to the overlapping of adjacent maxima or due to a change in the $\lambda_m$ of the peak that is being tracked by the spectrometer. An example of the latter is clear in Fig. 3a, where there are different $\lambda_m$ depending on the thermal range: for the range from 200 °C to 300 °C, $\lambda_m$ is around 1565 nm, whereas for the range from 400 °C to 500 °C it is around 1525 nm. This implies a requirement for data processing in order to obtain a coherent wavelength shift reading for the entire thermal range under study. This fact does not take place in the other two samples, as they have a well-defined peak which is the maximum at any temperature. However, between the aforementioned two samples, the sample in Fig. 3b shows the advantages of using a twin structure, which, as said before, imply higher visibility and narrower peaks. In this case, the same narrowness as the sample in Fig. 3c is achieved with half the length of fiber at the same time that only one peak is observed in the interrogation window compared to the three peaks in Fig. 3c.

In order to study the impact of $\Delta n$ on the thermal sensitivity, its variation with respect to the temperature was simulated by PhotonDesign for the two MCFs under study (see Fig. 4). In all cases, $\Delta n$ was evaluated at 1550 nm. To that end, the refractive index of the material was defined as $n = n_f + \gamma \Delta T$ [48], where $n_f$ is the refractive index of the cores at 25 °C, and $\gamma$ is the thermo-optic coefficient [39]. In this work, the non-linearity of the thermo-optic coefficient in the range from −25 °C to 900 °C has been taken into account for the simulations [49,50]. This is the reason why different slopes in the thermal range under study can be acknowledged in Fig. 4. As demonstrated in previously reported works [38], the expected proportional shift of $\lambda_m$ with temperature (as the temperature increases, $\lambda_m$ shifts to longer wavelengths, and vice versa) implies that the term $\left[ \frac{\partial \Delta n}{\partial T}*(1 + \alpha \Delta T) + \Delta n \alpha \right]$ in

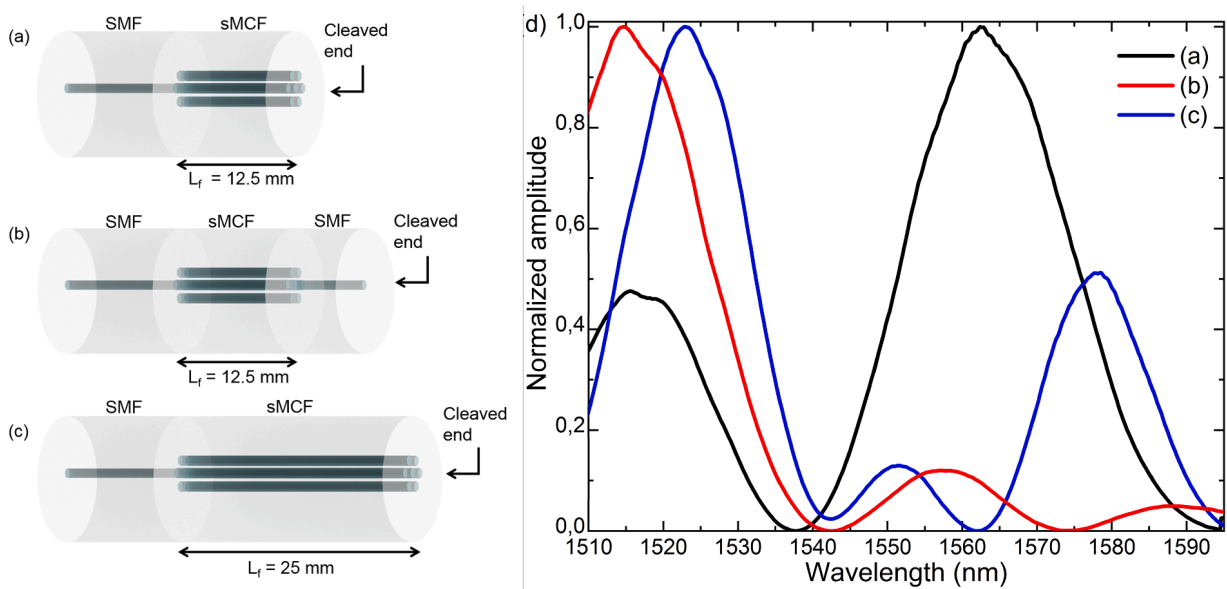

**Fig. 2.** (a)(b)(c) Scheme of manufactured samples used to demonstrate the negligible impact of the initial length of the MCF segments ($L_f$) on thermal sensitivity and (d) their corresponding spectra.

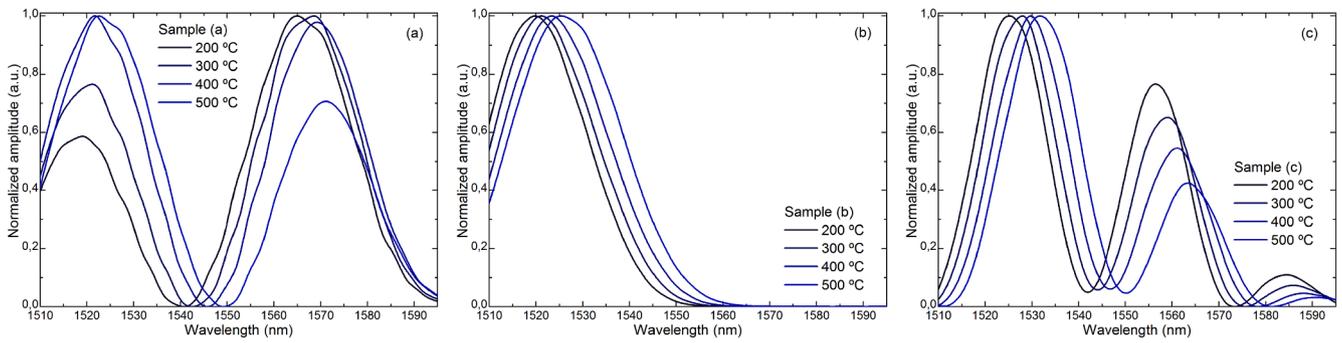

**Fig. 3.** Evolution of the spectra as a function of temperature of (a) sample (a), (b) sample (b) and (c) sample (c).

**Table 1**
Summary of the test.

| sMCF length ($L_f$) | 12.5 mm | 25 mm | Twin of 12.5 mm |
|---|---|---|---|
| Sensitivity (pm/°C) | 20.38 | 22.22 | 21.16 |

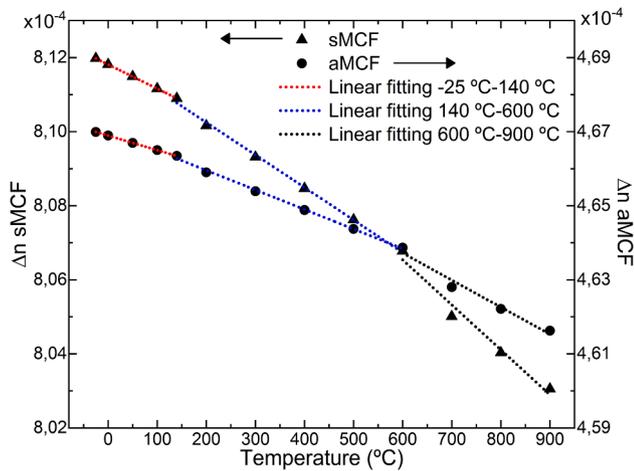

**Fig. 4.** Simulations of $\Delta n$ for the sMCF (black triangles) and aMCF (black circles). The black arrows indicate the vertical axis that corresponds to each fiber.

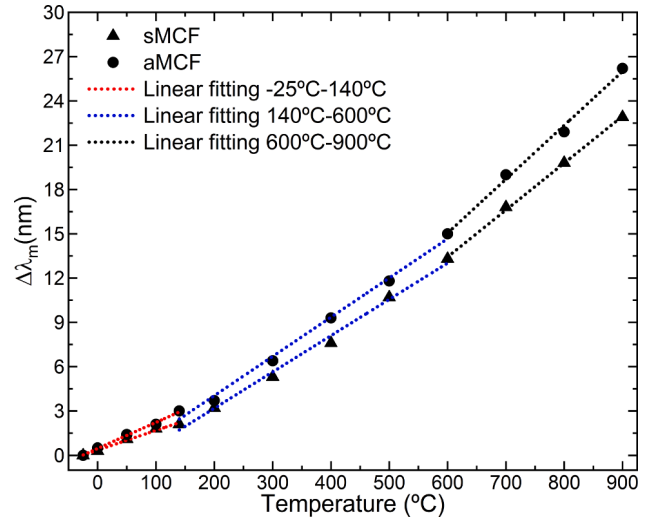

**Fig. 5.** Simulations of the expected wavelength shift for the sMCF (black triangles) and aMCF (black circles).

Eq. (9) is positive. From Fig. 4, it can be noticed that $\Delta n$ is always positive, but its slope (expressed as $\partial \Delta n / \partial T$) is negative in both MCFs under study. This indicates that $\left[\frac{\partial \Delta n}{\partial T} * (1 + \alpha \Delta T)\right] < \Delta n \alpha$, and as a result, it can be assumed that the thermal sensitivity increases as the difference between these two terms increases.

The conclusion from this theoretical analysis is that the thermal sensitivity of the aMCF is higher than that of the sMCF. This is because for the aMCF and in the temperature range under study, its $\Delta n$ is lower and its slope is smaller (see Fig. 4). In order to validate it, the expected wavelength shifts for segments of identical lengths of sMCF and aMCF were simulated (see Fig. 5). The information regarding the specific length of the fibers for this simulations is irrelevant as long as it is identical for both samples, as the aim was $\Delta n$ to be the only different parameter between both simulations. In Fig. 5, the higher sensitivity of the aMCF compared with that of the sMCF is noticeable, which is in good agreement with the theoretical conclusion.

To demonstrate experimentally the sensitivity enhancement of the aMCF compared to that with sMCF, a device of such aMCF was manufactured. For comparison purposes, the length of the aMCF (25 mm), its operating configuration in reflection mode and packaging were identical to the one reported in [38] so that the only difference between the two cases would be the $\Delta n$ of the different MCFs. In this manner, the effect of

$\Delta n$ could be analyzed and compared between fibers. The interrogation setup remained unaltered as for the previous tests. Two calibrated furnaces (Fluke 9150 and Fluke 9103) were used to cover a thermal range from −25 °C to 900 °C. The device was exposed to several stepped cycles from −25 °C to 140 °C with the Fluke 9103 and from 200 °C to 900 °C with the Fluke 9150 after being subjected to an annealing process to avoid hysteresis. The evolution of the spectrum as a function of temperature and the calibration curve of the manufactured device are shown in Fig. 6.

The manufactured device showed higher sensitivity than the one reported in [38] with the same packaging and operating configuration, reaching a sensitivity of 43.61 pm/°C for the range from 600 °C to 900 °C. These results are in good agreement with the theoretical values and conclusions extracted from Fig. 4 and with the simulation results in Fig. 5, which predicted the aMCF to be more sensitive than the sMCF for temperature measurements. Thus, it can be deduced that the most significant parameter to design highly sensitive MCF geometries is $\Delta n$.

By applying the conclusions regarding the effect of $L_f$ and $\Delta n$ for MCF-based thermometers, in this work, a sensitivity-optimized aMCF-based sensor for harsh environments was developed. For this purpose, the first step involved making the sensitive element (aMCF) as short as possible to make the sensor as compact as possible, enhancing its implementation ease to deploy it in real-field environments. However, shortening the MCF segment has its limitations, as $L_f$ does have an impact on the shape of the spectrum, as demonstrated in Fig. 2d and in Fig. 3. According to Eq. (2), as the MCF segment is longer, the period of the spectrum is shorter, and as a result, narrower peaks are obtained, which implies more peaks in the interrogation window of the spectrometer. If the distance between adjacent peaks in the spectrum is shorter than the shift caused by temperature, an ambiguity in the measurements will take place. To avoid that, a compromise is necessary regarding the length of the MCF so that the period of the spectrum is longer that the expected shift in the measured temperature range. To that end, simulations with the PhotonDesign were carried out to obtain the shortest $L_f$ of the aMCF that suited this requirement. From the simulation results, the best fitting aMCF length was found to be 12 mm. Fig. 7 shows both manufactured and simulated spectra, where it can be noticed that the manufactured spectrum fits with the simulated one.

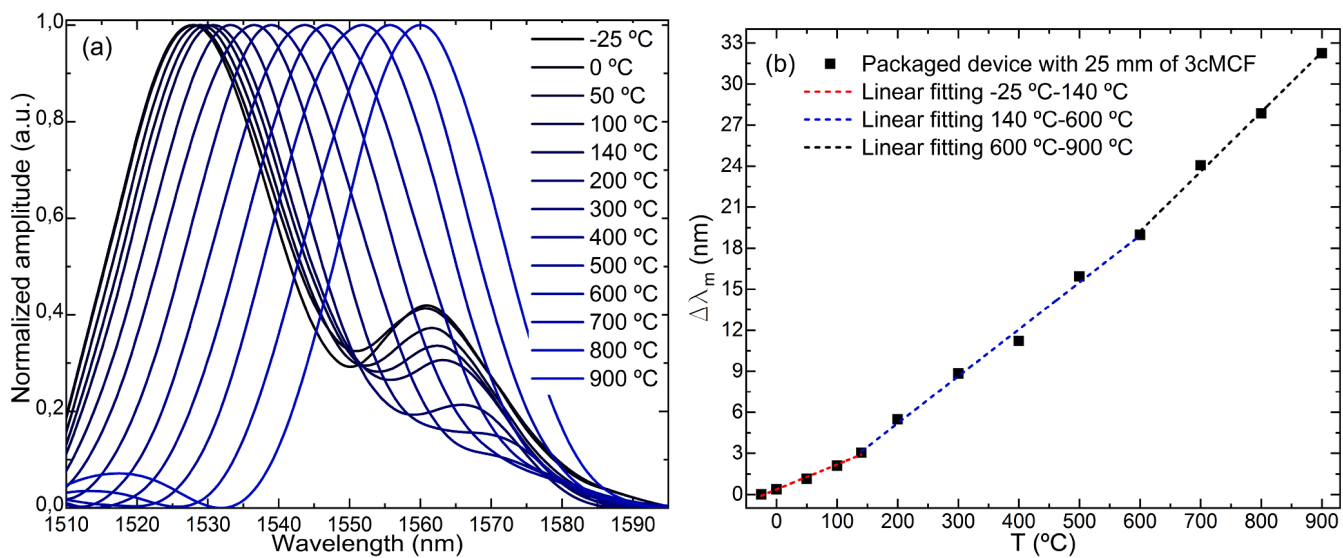

**Fig. 6.** (a) Evolution of the spectrum as a function of temperature and (b) calibration curve of the manufactured device with 25 mm of aMCF.

The second step involves a packaging capable to withstand extreme environments. In the device whose performance is shown in Fig. 6, the thin stainless-steel metallic shielding tended to bend when it was exposed to extreme temperatures and did not return to its original shape, which is a potential problem to operate in harsh environments as it may lead to break the fiber inside it (see Fig. 8). To overcome such a critical drawback, a ruggedized packaging was developed in order to make it robust to operate in harsh environments and extreme temperatures. The packaging of this device consisted of three layers that covered the optical fiber. The first layer was the same as for previous devices and consisted of a ceramic tube (Omega TRX-005132–6) to keep the MCF straight and therefore, make it only sensitive to temperature. The second layer consisted of a thin Inconel tube (INC-116–6-OPEN) whose inner diameter is slightly bigger than the outer diameter of the ceramic tube. Its aim was to provide rigidity to the ceramic tube and avoid any possible fracture of it, as the latter shows fragility against impacts. The third layer consisted of a thicker Inconel tube (INC-18-6CLOSED) to provide robustness. The reason for using Inconel was due to the fact that it is capable to withstand higher temperatures with high structural integrity. With this ruggedized packaging, the final prototype was about 14 cm long (see Fig. 8).

Although the sensing part of the device was only 12 mm long and located at the tip of it, the 14 cm-long ruggedized packaging was caused by the configuration of the furnaces in which the device was tested. They have a circular hole where the sensor needs to be inserted vertically and that only at its deepest point reach their operating temperature. Therefore, to protect the parts of the sensor that are not sensitive to temperature but are inside the furnace, such packaging is required. Finally, to make the device as robust as possible to be deployed in extremely harsh environments, the SMF from the device to the interrogation setup was protected as well with a double layer. The first one consisted of a tube (Thorlabs FT030) that contained an outer PVC jacket and Kevlar threads to provide protection. The second layer consisted of a stainless steel jacket (Thorlabs FT05SS) which provided extra protection and avoided any visible or IR light entering through the length of the fiber.

## 3. Results and discussion

The interrogation setup of the aMCF-based ruggedized device was identical to the one used and explained throughout this document, and, as it has been done with all the devices in this work, before running the tests, it was subjected to an annealing process to avoid hysteresis. After that, the device was exposed to several stepped temperature cycles from −25 °C to 140 °C with the Fluke 9103 and from 200 °C to 900 °C with the Fluke 9150. The position of the maximum peak ($\lambda_m$) in the spectrum as a function of time of one of the performed cycles of each of the two tested temperature ranges is shown in Fig. 9.

From Fig. 9, it can be noticed that the duration of the steps of the cycles for the range from −25 °C to 140 °C are shorter than those for the range from 200 °C to 900 °C. The reason for that is that the temperature changes are smaller and closer to room temperature. Therefore, the furnace Fluke 9103 requires less time to reach and stabilize in such values than the Fluke 9150 for higher temperatures. For the range from −25 °C to 140 °C, each step lasted 60 min and each of the cycles like the one in Fig. 9a lasted 9 h. For the range from 200 °C to 900 °C, each step lasted 90 min and each of the cycles like the one in Fig. 9b lasted 22.5 h.

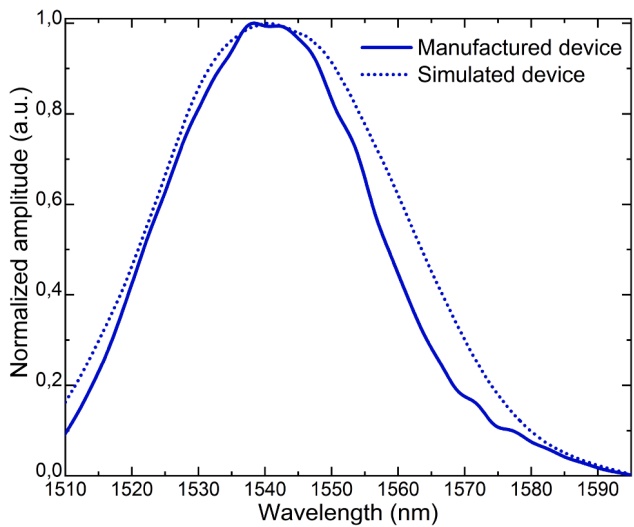

**Fig. 7.** Spectra of the manufactured and simulated devices.

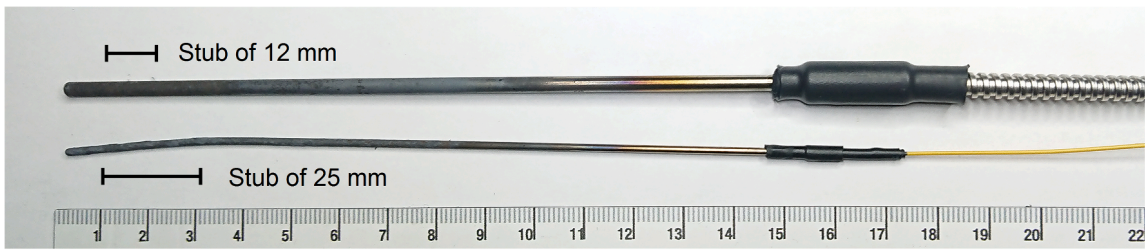

**Fig. 8.** aMCF-based manufactured devices. The sensing elements are located far enough from the tip of the device to avoid creating a Fabry-Perot cavity. The blackening gradient indicates that the highest temperature has been reached where the MCF segments are located.

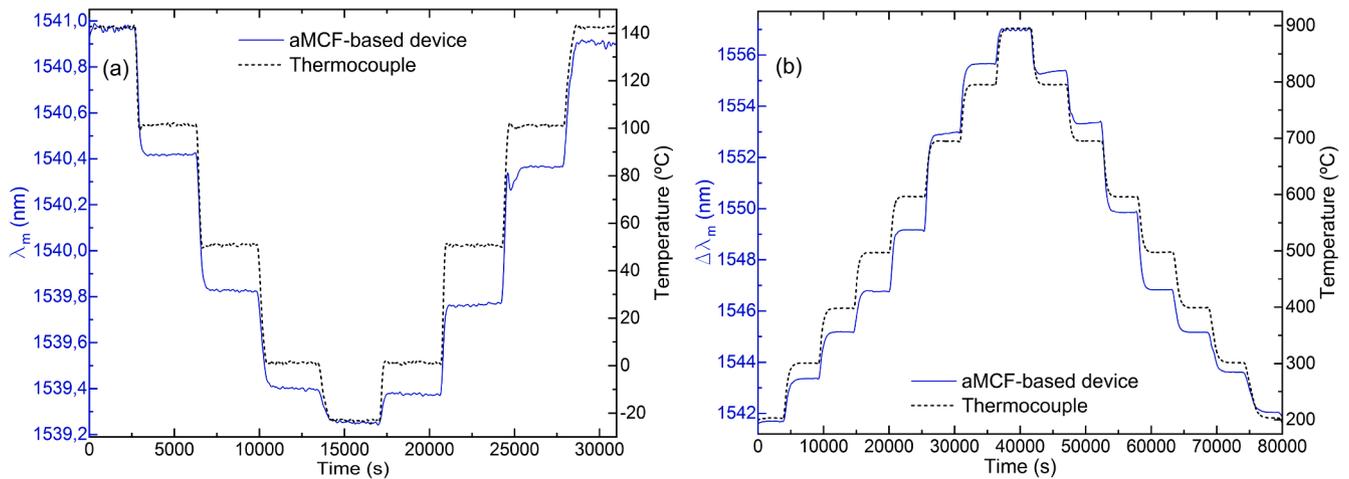

**Fig. 9.** Time evolution of our packaged aMCF sensor compared to that of the thermocouple for the ranges (a) from −25 °C to 140 °C and (b) from 200 °C to 900 °C.

The evolution of the reflection spectra of the aMCF-based device as a function of temperature is shown in Fig. 10a.

The proportional increase in sensitivity with temperature is easily noticeable. This behaviour was expected from the conclusions extracted from Fig. 4 and Fig. 5. Additionally, the device shows a unique and well defined peak during the entire operation range, as it was pursued in its design.

For comparison purposes, an identical device was manufactured with the same physical characteristics, dimensions and packaging but with sMCF. This device was subjected to the same annealing and calibration processes as the ruggedized aMCF-based device. The evolution of its spectrum as a function of temperature is shown in Fig. 10b.

An increase in sensitivity proportional to temperature can be acknowledged as well, although such increase is significantly smaller than that in the aMCF-based device.

The calibration curves for the ruggedized sensors are shown in Fig. 11. Results demonstrate that the thermal sensitivity of the aMCF is higher than the identical device with sMCF in all the tested temperature ranges, obtaining almost twice the sensitivity in the range from 200 °C to 900 °C. For the three expected and acknowledged temperature ranges (-25 °C-140 °C, 200 °C-600 °C and 600 °C-900 °C), the sensors showed linear behaviors with different sensitivities, due to the aforementioned non-linear nature of the thermo-optic coefficient. Thus, there will be a unique and specific correlation between temperature and the tracked peak ($\lambda_m$) for each of these sectors. Depending on the expected thermal operating range, the corresponding correlation will have to be selected to link wavelength measurements (in nm) with temperature (in °C). The results are summarized in Table 2., which agree with the fact that as $\Delta n$ decreases, the sensitivity increases proportionally, as it was deduced in the theoretical analysis discussed above.

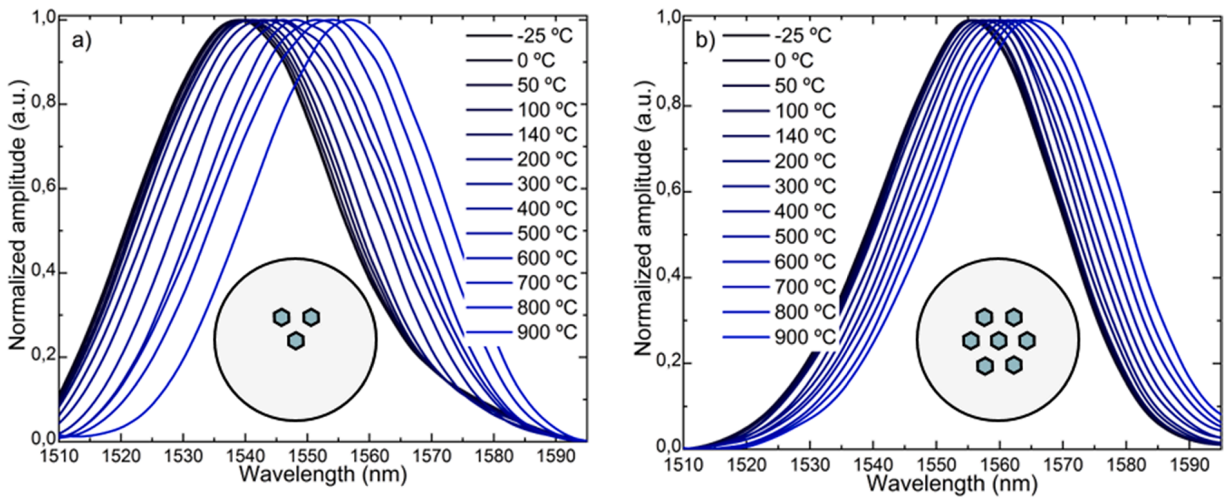

**Fig. 10.** Evolution of the spectra of the (a) aMCF and (b) sMCF-based ruggedized device as a function of temperature.

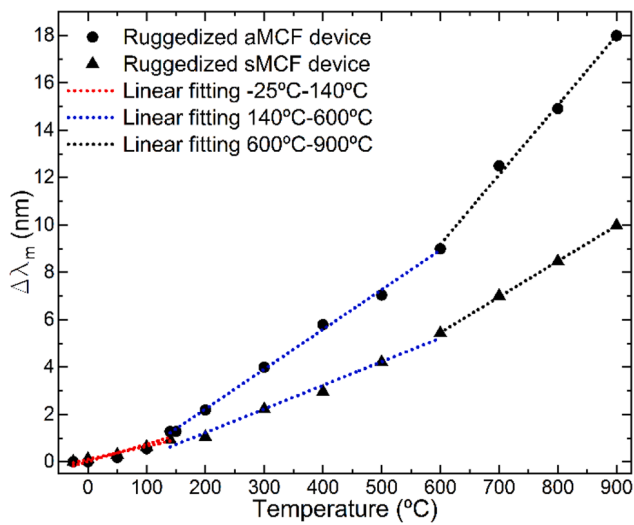

**Fig. 11.** Calibration curves for the ruggedized aMCF device (black circles) and the ruggedized sMCF device (black triangles).

Moreover, from comparing the results for the aMCF-based devices in Fig. 6 and Fig. 10a, it can be noticed that the sensitivities have identical trends although with different values. As the only difference between such cases is the packaging, results suggest that the latter is responsible for this loss in sensitivity, as the temperature transference decreases from the environment to the inside of the packaging and to the MCF. Identical conclusion can be obtained from the results of the device made of sMCF if we compare its results in Fig. 10b with the ones in [38] with a less robust packaging. Our results demonstrate clearly that by selecting adequately the MCF, the device may be ruggedized to make it capable to withstand extremely harsh environments at the same time that it is still more sensitive (29.426 pm/°C) than previously reported solutions [38].

## 4. Conclusion

In this work, we have reported on a ruggedized optical thermometer based on a strongly coupled MCF capable to withstand extremely demanding working conditions. It consists of a segment of aMCF (12 mm) fusion spliced to the distal end of a standard SMF. In order to operate in reflection mode, the end of the MCF was cleaved so it acted as a low reflectivity mirror. The sensor was ruggedized with a triple layer shielding consisting of an inner ceramic tube and two outer Inconel tubes. In this manner, the MCF was kept straight to avoid any bending or curvature to make it sensitive only to temperature, and protected against impacts.

The sensor was calibrated for a thermal range from −25 °C to 900 °C and a K-type thermocouple was used as reference to obtain the correlation between the wavelength shift in the spectrum and the temperature. Results show that the device has a sensitivity up to 29.426 pm/°C, which is higher than previously reported packaged MCF-based thermometers with identical SMF-MCF configuration and more fragile shieldings. To the author's best knowledge, this is the first time where such combination of high sensitivity, ruggedized packaging and wide thermal operating range is reported for MCF-based thermometers.

The results reported here highlight the significance of understanding the fundamentals of strongly coupled MCFs for temperature sensing, as thanks to it and as it has been demonstrated in this work, the most thermal sensitive MCF among several geometries can be predicted with high accuracy. As a result, it allows a more robust shielding, as the loss in sensitivity due to the ruggedized packaging is compensated by the increase in sensitivity provided by the appropriate selection of the MCF. Moreover, the fundamentals discussed here open the possibility to design and manufacture ad-hoc MCF structures and geometries with optimized and/or enhanced sensitivities to measure temperature or any other parameter of interest such as strain or bending, or to design filters or switches, as the procedure for all those cases is analogous to the one exposed herein.

We believe that the sensor reported here is an attractive alternative for many different applications that require temperature measurements in very demanding environments, as it is ruggedized, capable to operate in a wide thermal range and highly sensitive, especially for temperatures above 600 °C. The sensor is compact, simple, and easy to fabricate and interrogate, which makes it cost-effective and reproducible. Therefore, we believe this prototype is a step forward towards commercially viable optical thermometers.

## Declaration of Competing Interest

The authors declare that they have no known competing financial interests or personal relationships that could have appeared to influence the work reported in this paper.

## Acknowledgments

Ministerio de Economía y Competitividad; Ministerio de Ciencia, Innovación y Universidades; European Regional Development Fund (PGC2018-101997-B-I00 and RTI2018-094669-B-C31); Gobierno Vasco/Eusko Jaurlaritza (IT933-16); ELKARTEK KK-2019/00101 (µ4Indust) and ELKARTEK KK-2019/00051 (SMARTRESNAK). The work of Josu Amorebieta is funded by a PhD fellowship from the University of the Basque Country UPV/EHU. The work of Angel Ortega-Gomez is funded by a PhD fellowship from the MINECO (Ministerio de Economía y Empresa de España).

**Table 2**
Summary of the results of the ruggedized devices.

|  |  | −25 °C-140 °C | 200 °C-600 °C | 600 °C-900 °C |
|---|---|---|---|---|
| aMCF | Correlation | T = 117.61 $\lambda_m$ + 5.74 | T = 59.445 $\lambda_m$ + 67.693 | T = 33.83 $\lambda_m$ + 290.09 |
|  | Sensitivity (pm/°C) | 7.302 | 16.785 | 29.426 |
|  | R² | 0.926 | 0.9989 | 0.997 |
|  | $\sigma^2$ (nm²) | 0.81173 | 0.35215 | 0.4407 |
| sMCF | Correlation | T = 177.31$\lambda_m$- 17.193 | T = 86.72 $\lambda_m$ + 129.17 | T = 66.295 $\lambda_m$ + 238.47 |
|  | Sensitivity (pm/°C) | 5.531 | 9.983 | 15.081 |
|  | R² | 0.99 | 0.992 | 0.999 |
|  | $\sigma^2$ (nm²) | 0.44623 | 0.61978 | 0.14858 |